\documentclass[conference]{IEEEtran}
\usepackage{cite}
\usepackage{amsmath,amssymb,amsfonts}
\usepackage{algorithmic}
\usepackage{graphicx}
\usepackage{makecell} 
\def\BibTeX{{\rm B\kern-.05em{\sc i\kern-.025em b}\kern-.08em
    T\kern-.1667em\lower.7ex\hbox{E}\kern-.125emX}}
\begin{document}

\title{
Enabling Fast Exploration and Validation of Thermal Dissipation Requirements for Heterogeneous SoCs
}


\author{\IEEEauthorblockN{Joel Öhrling, Dragos Truscan, and Sebastien Lafond}
\IEEEauthorblockA{\textit{Department of Information Technology} \\
\textit{Åbo Akademi University}\\
Turku, Finland\\
joelohrling@gmail.com, dragos.truscan@abo.fi and sebastien.lafond@abo.fi}
}

\maketitle

\thispagestyle{plain}
\pagestyle{plain}

\begin{abstract}

 The management of the energy consumption and thermal dissipation of multi-core heterogeneous platforms is becoming increasingly important as it can have direct impact on the platform performance.  This paper discusses an approach that enables fast exploration and validation of heterogeneous system on chips (SoCs) platform configurations with respect to their thermal dissipation. Such platforms can be configured to find the optimal trade-off between performance and power consumption. This directly reflects in the head dissipation of the platform, which when increases over a given threshold will actually decrease the performance of the platform. Therefore, it is important to be able to quickly probe and explore different configurations and identify the most suitable one. However, this task is hindered by the large space of possible configurations of such platforms and by the time required to benchmark each configurations. As such, we propose an approach  in which we construct a model of the thermal dissipation of a given platform using a system identification methods and then we use this model to explore and validate different configurations. The approach allows us to decrease the exploration time with several orders of magnitude. We exemplify the approach on an Odroid-XU4 board featuring an Exynos 5422 SoC. 
 

\end{abstract}

\begin{IEEEkeywords}
thermal dissipation, heterogeneous SoC, system identification, thermal profiling, thermal prediction
\end{IEEEkeywords}

\section{Introduction}



In the age of ever-increasing demand for mobile data traffic and the number of Internet-connected devices growing every day, new technologies to meet these demands on throughput and availability are continuously being researched and deployed. With ever more dire climate reports presented every year, it is also crucial to consider the climate impact of these telecommunications systems.
In 2016, it was estimated that around 5\% of the world's CO$_2$ emissions originated from information and communication technology \cite{7446253}. The same year YouTube alone contributed with 10 million tons of CO$_2$-equivalent emissions \cite{Preist:2019:ESI:3290605.3300627}, roughly twice the annual carbon footprint of the Helsinki Metropolitan Area \cite{ilmastotekoja-2018}.

Over the past 15 years, however, there has been a shift towards increasing the number of cores instead of the clock frequency~\cite{rupp_2018}. Today, there exists a wide range of chips that integrate several types of components into the same casing.  These are referred to as a System-on-a-Chip (SoC). An SoC is typically a fully functioning computer on a single chip. Components that have traditionally been discrete parts in a large computing system are now being integrated on the same chip. For heterogeneous systems, this means that multiple types of processing units are integrated into the same chip. These types of chips are generally referred to as heterogeneous SoCs or heterogeneous processors. Heterogeneous processors can, for example, be found in most of today's mobile phones and are becoming progressively more prevalent in all types of computers \cite{mcguinness_2013}. This is also the case for information and communications systems \cite{ullman_2013}. The primary motivation for deploying heterogeneous processors in these systems is to provide better efficiency, as different types of cores are optimized for different types of instructions or workloads \cite{wolf_1041_nodate}.


A heterogeneous SoCs is a standardized specification that allows for the integration of different types of processors on the same bus. Architectures that combine two or more multi-core CPUs are conventional in many energy-constrained devices today. The purpose of these architectures is to provide better energy efficiency by combining cores that have different power consumption characteristics and performance.

With the increase of heterogeneous SoCs, extra functional properties (EFP) of cyber-physical systems (CPS), like energy consumption and thermal dissipation are becoming increasingly important. In addition to a direct increase in the associated cooling costs, an increase of energy consumption and thermal dissipation above platform specific thresholds can result in degraded application performance which in turn can reflect in financial and reputation losses for the service provider. 


Testing and verification of the thermal characteristics of a heterogeneous multi-core processor can be an exhaustive process with many parameters to consider. Even for a single-core processor, there is a large number of variables that impact how much heat a processor generates, for example, the ambient temperature, the workload application and the core frequency. For heterogeneous multi-core platforms, this number grows even further as each processing element comes with its own unique characteristics. Performing exhaustive testing on these types of systems becomes unfeasible, as the number of possible configurations is vast. 

For example, standard use cases could require several years of testing time to exhaustively explore all possible configuration of a eight cores ARM big.LITTLE architecture.
Constructing models of a system is, therefore, an alternative that can be considered. Building a model can enable the prediction of thermal dissipation in a system and can help speed up the test and verification process.
The need for constructing accurate models for the validation of extra-functional properties of the cyber physical systems has been advocated by many authors~\cite{Yuan2019}. Using explicit models of the system is not always feasible as such models require detailed low-level knowledge of the system implementation. Such information is in most cases only available to the silicon manufacturers. 


Therefore in this paper, we provide an approach for fast exploration and validation of thermal dissipation requirements for heterogeneous SoCs which takes advantage of system identification  methods to create a thermal model of the system. The obtained model can be used not only for fast exhaustive exploration for platform configurations, but also for validating, for example at runtime, that different selected configurations satisfy the thermal dissipation requirements of the system. We exemplify the proposed approach on image processing application executed on a Exynos  5422  SoC, and demonstrate that an approach relying on the state-space system identification algorithm can produce accurate prediction in terms of heat generation and temperature of the processor die.

\section{Motivating example and proposed approach}
\begin{figure}[h!]
\centering
  \includegraphics[width=0.92\linewidth]{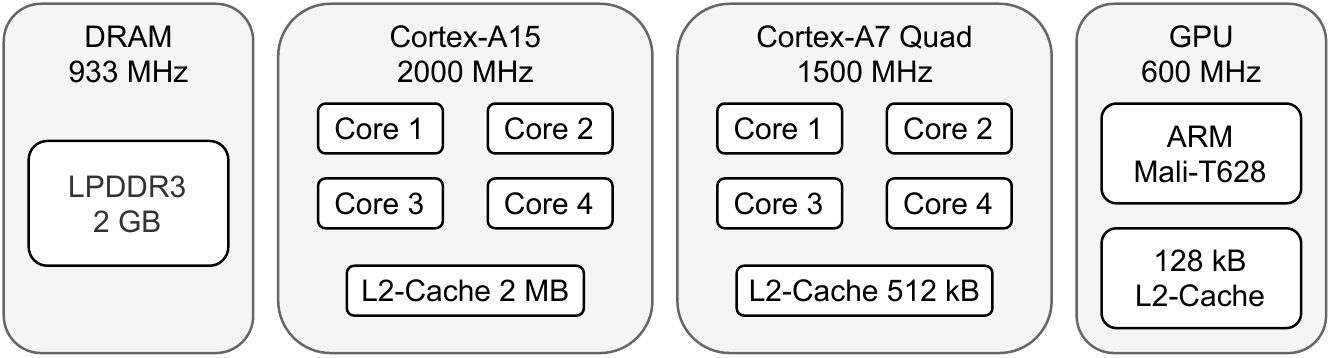}
  \caption{Internal block diagram of the Exynos 5422 SoC.}
  \label{fig:block}
\end{figure}

As a motivating example, let us consider an Odroid-XU4 board manufactured by Samsung and featuring an Exynos 5422 SoC and 2 GB of DDR3 memory.  The block diagram of the Exynos processor is shown in Figure \ref{fig:block}. Even if it is a quite simple board, it is a good representative example of a heterogeneous SoCs platform. The board has eight cores configured in two clusters and runs the Ubuntu operating system based on the Linux 4.14 kernel. The Exynos 5422 implements the big.LITTLE heterogeneous computing architecture developed by ARM. In this architecture, one cluster is more potent in terms of computing power, but also thirstier in terms of power dissipation. The other cluster is smaller and has less computing power, while being more energy-efficient. In this heterogeneous processor, an ARM Cortex-A15 is implemented as the big cluster and an ARM Cortex-A7 is implemented as the little cluster.

Different configurations of the board can be created by changing parameters such as CPU and GPU cluster frequency, utilization level for each core, etc. in order to obtain the best trade off between power consumption and the performance of the platform. However, the large amount of parameters and their ranges result in a very large configuration space, which will make the exploration and selection of valid configurations difficult.

For instance, the cluster frequency for each of the two CPU clusters on the Odroid-XU4 can range between 200MHz  and 1500MHz  for  the  little cluster and  between 200MHz  and 2000MHz  for  the  big cluster, and the utilization threshold can be varied between 0 and 100\%. This will result in 234 * 10\textsuperscript{6} combinations.

This is even more time consuming when evaluating the head dissipation for a given configuration. On this particular board, and depending on the class of application being run, the steady state temperature of a given configuration is reached on average after 100 seconds. If one were to explore all previous configurations with respect to head dissipation of each, it will require 234*10\textsuperscript{8} seconds,  which represent roughly 742 years. Of course, different optimizations can be applied in order to reduce the number of combinations, e.g., sampling the processor frequencies in 100 MHz increments, or removing some already known infeasible parameter configurations, still the configuration space remains impractical to explore. 

\begin{figure*}[t!]
\centerline{\includegraphics[width=0.75\textwidth]{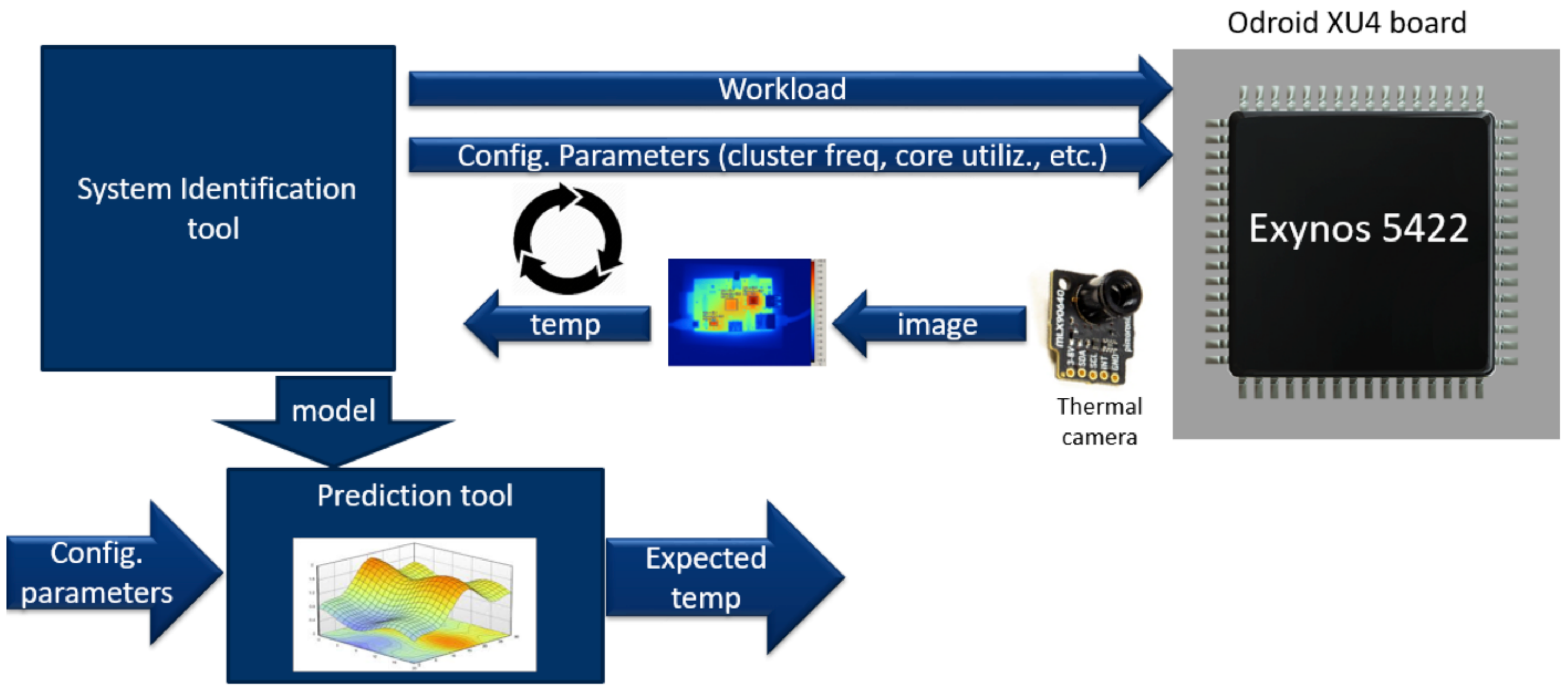}}
\caption{Overview of the experimental setup}
\label{overview}
\end{figure*}

Therefore, in this paper we propose an approach for fast exploration and validation of thermal dissipation requirements for heterogeneous SoCs as shown in Figure \ref{overview}. The approach relies on the system identification theory to build a model of the SoC under evaluation by performing a series of external (black box) observations  that covers only a fraction of the configuration space of the platform.  The created model is later on used to explore exhaustively entire configuration space in search of best configurations for a given platform and application or to answer questions such as \textit{"what will be the generated heat of this heterogeneous CPU platform (HCP) for a particular configuration}". The answer to the above questions would allow not only find the best configuration but also to validate that the system satisfies its thermal requirements. 

\section{Preliminaries}

\subsection{Power and heat generation}

Although heterogeneous SoCs introduce many new possibilities for the power and thermal management of computers, the same principles that dictate the power and thermal dissipation of a single-core CPU still applies.

The power that a processor dissipates originates from the large number of transistors that are switching as the CPU executes instructions. The total power consumption of a CPU core can be described as the sum of three types of power consumption: the dynamic power consumption, the static power consumption and the short-circuit power consumption. The short-circuit power has, however, been shown to be practically negligible in modern CPUs \cite{7132972}. Therefore, Equation \eqref{eq:dvfs} is the formula commonly used to describe the power consumption, where $P$ is the total power, $P_{dyn}$ is the dynamic power and $P_{sta}$ is the static power.
\begin{equation}
\label{eq:dvfs}
P = P_{dyn} + P_{sta}
\end{equation}

The heat produced by a CPU is closely coupled to its power dissipation. As shown by the conduction equation \eqref{eq:fouriers_law} and the convection equation \eqref{eq:newton1}, the amount of heat that is transferred from a processor to the ambient is dependent on the temperature difference between the processor and the ambient. 

\begin{equation}
\label{eq:fouriers_law}
Q = - kA \frac{dT}{dx}
\end{equation}

\begin{equation}
\label{eq:newton1}
Q = hA \frac{dT}{dx}
\end{equation}

with $Q$ is the heat transfer rate $(W)$, $k$ is the thermal conductivity of the material $(W / m K)$, $A$ is the cross-section area in the heat flow direction $(m^2)$ and $\frac{dT}{dx}$ is the temperature gradient $(K/m)$ in the direction of heat flow. Conduction is the main source of heat transfer inside a processor die \cite{golicha}.  $h$ is a heat transfer coefficient that depends on many factors, such as the geometry of the object's surface and the fluid's viscosity. Inside a processor, convection does generally not occur, as there are no fluid components. Convection is, however, the main source of heat dissipation between the processor and the ambient environment, usually through the air \cite{MORADIKAZEROUNI20191218}.

The heat produced by a CPU is closely coupled to its power dissipation. As shown in Equation \eqref{eq:fouriers_law} and \eqref{eq:newton1}, the amount of heat that is transferred from a processor to the ambient is dependent on the temperature difference between the processor and the ambient.
Hence, the heat dissipation can be viewed as a dynamic system that can be represented by the following differential equation:

\begin{equation}
\label{eq:tempdiff}
C\frac{dT}{dt} + \frac{T-T_{amb}}{R} = P
\end{equation}

Here, $T_{amb}$ is the ambient temperature and $R$ and $C$ represent the thermal resistance and conductivity of the chip, respectively. Based on this, the thermal-electric analogy can be utilized to model the temperature dynamics of a processor as an nonlinear RC-circuit.

\subsection{System Identification}

The area of system identification (SI) \cite{lennart1999system} is a research domain which creates mathematical models of dynamical systems through statistical and machine learning approaches. System identification also deals with how to model a system using limited amounts of data. 

In SI, a \textit{system} is a conceptualization of a real-world process, such as a physical process or the mechanisms of economics on the stock market, while a \textit{model} is a relation between measured quantities. Most commonly, this relation expresses how one or more inputs map to one or more outputs. The relationship is typically expressed as a mathematical formula, but it can be any type of function, such as a lookup table. A model is, thus, a manageable representation of a system that seeks to approximate the system.  The difference between a model and the system the model represents is commonly known as the \textit{approximation error} or prediction error and it reflects in the difference between the output of the actual system and the approximation of the system that the model represents.
    
A \textit{regressor} is an independent variable used as input when estimating a model. A regressor can also be known as a feature, a term that is commonplace in machine learning contexts. A \textit{regressand} is a dependent variable that the regressors are used to predict the outcome of.
    
In data-driven modeling and machine learning, training, validation and test sets are commonly utilized. The training set is the data the model learns from and tries to mimic. The validation set is used to tune the hyperparameters of the model. \textit{Hyperparameters} represent parameters that are not learned. These are often factors that impact the model structure itself or how it is trained. The test set is utilized to provide an unbiased evaluation of the model's ability to generalize to unseen data. The model's performance is generally assessed using the test set. The test set is sometimes also known as the holdout set.

In system identification, modeling approaches are usually categorized into one of three groups: white-box modeling, gray-box modeling and black-box modeling. White-box modeling is an entirely theoretical modeling approach, where the model is constructed based on knowledge about the system and does not rely on data. White-box approaches require a complete knowledge of all the properties, parameters and uncertainties of a system. Black-box modeling is the complete contrast to white-box modeling. Black-box approaches depend solely on observations of a system without relying on insights into the underlying process. Data-driven modeling is another name to describe black-box modeling. The third group of system identification techniques is gray-box modeling. Gray-box techniques rely on parts of both black- and white-box modeling.

\section{Related work}
Many approaches have been proposed in the field of thermal modeling from theoretical white-box models to black-box models using neural networks. A common approach to thermal modeling is to utilize power consumption measurements from the processor and exploit the relationship between power and thermal dissipation. Heterogeneous SoCs are, however, not commonly equipped with sensors measuring the power consumption for each individual core. Therefore, our work will consider modeling approaches that strictly rely on the core frequencies and each core's utilization percentage to make predictions about thermal dissipation.  This section presents a review of previous research that has been performed in the fields of nonlinear system identification and thermal modeling of processors.  

Approaches towards the black-box end of the system identification spectrum that utilize neural networks have  been proposed, for instance, by  Vincenzi et al. \cite{5993628} and Sridhar et al. \cite{6106739}. Both have suggested two similar implementations where the thermal dynamics of an integrated circuit is predicted using ARX linear neural networks. These approaches were shown to be effective at simulating heat flow in 3D-dimensional and highly granular, integrated circuits. An approach to simulating the heat dissipation in processors using a feed-forward neural network has also been proposed in \cite{zhang_machine_2018}. The researchers compared the performance of a Gaussian process model, a neural network (NN) model and a linear regression model. The results showed that the neural network model outperformed the linear model in terms of prediction accuracy by 30\%, but was approximately three times as computationally expensive. The Gaussian process model also showed good prediction accuracy. However, it had twice as much computational overhead as the NN model. 

Another interesting approach was tested by Pérez et al. \cite{perez} in an article from 2018. They compared recurrent and feedforward neural network structures for thermal prediction of immersive cooling computer systems. A simple feedforward ANN structure was compared with two other structures utilizing LSTM and GRU layers in an FIR configuration. The temperature predictions in this study were based on the core frequency and processor utilization measurements from the past minute. The results revealed that the shallow GRU and LSTM structures produced the lowest prediction error.

Many of the approaches to thermal modeling presented in the previous section rely on power measurements to predict the temperature of a processor. There are examples in the literature of research that proposes methods for predicting the power dissipation of a processor. Walker et al. \cite{walker0,walker1,walker2} utilize core frequencies, core voltages and event counters to train a linear regression model to predict the power consumption of a multi-core processor. The events used were, for example, cycle counter, bus and cache accesses. Zhang et al. \cite{zhang} implement a similar approach. They also build a linear regression model based on data collected from a CPU. Unlike, Walker et al, however, they utilize the idle states and idle time of each core. Another similar approach has been suggested by Balsini et al \cite{balsini}. The latter approach deploys a genetic algorithm to find the optimal parameters for a function that represents the theoretical relationship between power dissipation and quantities such as the core voltage and clock frequency. 

A neural network-based approach to modeling the power of CPU has been proposed in \cite{djedidi:hal-01856579}. In this paper, Djedidi et al. construct a power model for an ARM-based mobile device using a NARX structure. The model is constructed based on data such as core frequency, screen activity and network usage. 

The works above presented in this section reveal that a wide range of methods has been applied to model or simulate temperature development in processors, computing systems and adjacent areas. Most approaches have relied on more classical gray-box system identification techniques. However, articles published in recent years have focused more on machine learning techniques and especially neural network approaches. 

Using neural networks for exploration of heterogeneous SoCs has also been proposed in \cite{Ivan9155898}. There, the authors use a discriminator network to efficiently find configurations that satisfy processing requirements and power consummation constraints by exploring only a fraction of the configuration space. The approach was also easy to integrate in continuous integration frameworks. However, the approach requires an on-the-fly exploration, which in the case of thermal dissipation measurements,  will be heavily influenced by the duration of each temperature measurement to reach a steady state.

\section{Experimental setup}

\begin{figure}[h!]
\centering
  \includegraphics[width=0.8\linewidth]{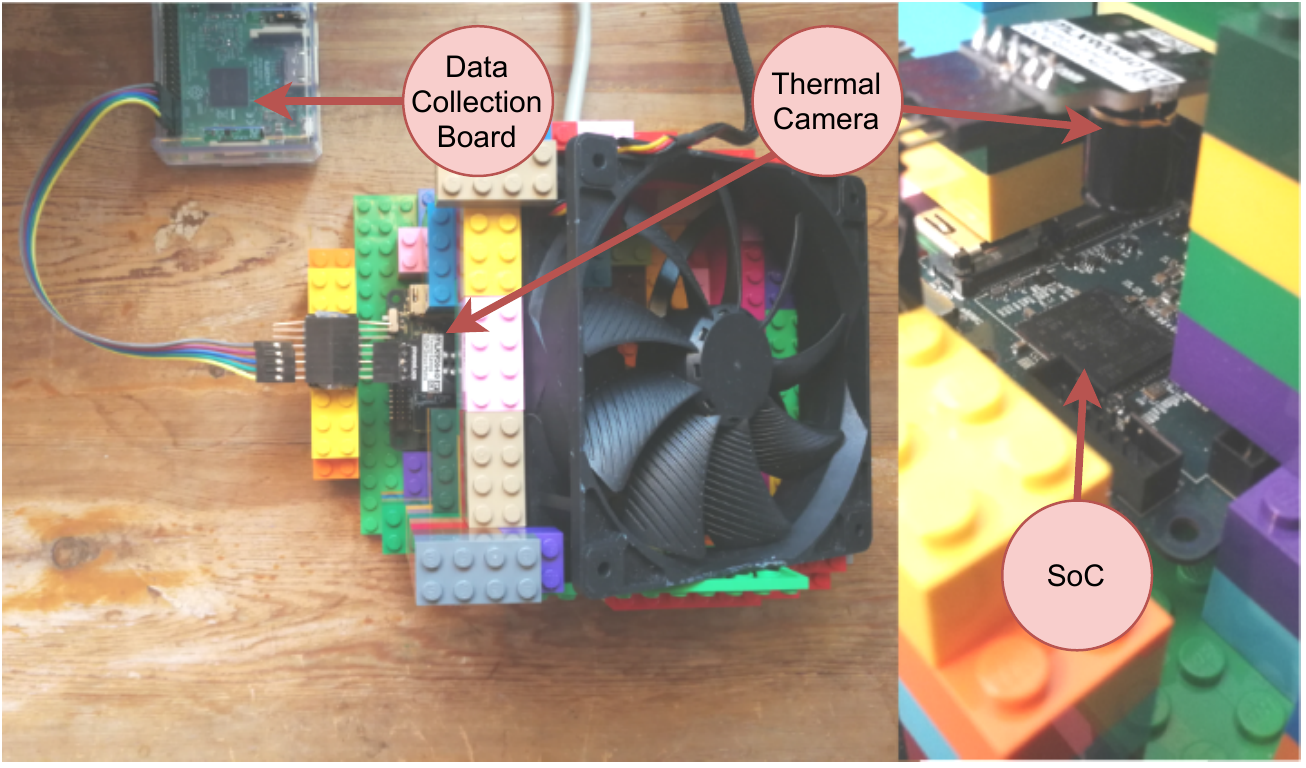}
  \caption{The experimental setup.}
  \label{fig:setup}
\end{figure}

\label{section:CaseStudy}

For this study, a desktop experiment setup for bench-marking and measuring the temperature of a heterogeneous processor was created. Following is a description of all parts in the experimental setup. Figure \ref{fig:setup} shows the entire setup that was used to generate and gather data in this study.  The setup features four parts: the heterogeneous SoC that is the system under test, a thermal sensor, a cooling fan, and a data collection unit. The remainder of this section will detail the parts of this setup and how data was collected from the heterogeneous processor.

This single-board computer allows for control of the frequency on a per cluster basis between $200$ MHz and $2000$ MHz for the big cluster and $200$ MHz and $1500$ MHz for this little cluster. The operating frequency cannot be controlled independently for each core inside the clusters. The voltage levels can also be set for each cluster. However, in the Linux operating system for this platform, these are set to static values for each operating frequency by the kernel. The operating voltage levels are, therefore, not considered as a variable in the implementations in this work.

The Odroid board has been configured to trigger a thermal throttle when the core temperature for the big cores reaches 90$^{\circ}$C. This means that the processor's frequency governors will reduce the maximum available frequency when the temperature is reached to prevent the processor from overheating.

\subsubsection{Experimental workload} \label{sec:workload}
The workload application utilized in this project is an RGB-YUV image conversion. This image conversion was chosen as the workload because it is a highly parallel workload that can be distributed to many cores. The workload implementation was taken from the \textit{stress-ng} bench-marking program \cite{stress-ng}. This program, however, could not be utilized on its own, as it did not feature the ability to control the utilization on a per core basis. Furthermore, the core frequencies could not be adjusted by this stress testing suite. Thus, a custom-built stress application has been implemented in this work.

The stress application is written in C. It implements the previously described workload. The application takes the utilization of each core and the frequency of each cluster as well as the amount of time it should be executing as arguments. Inside the application, a thread for each core in the system is created. Each core thread runs its assigned workload independently from the other cores. 

The software manages the utilization control separately for each core's thread. This is achieved by allotting periods of 10 milliseconds. For each period, work is performed for the specified core utilization. The core thread is then put to sleep for the remainder of the period using the \textit{select} system call. The \textit{select} call allows the processor to sleep without deallocating any of its resources. This method of utilization control requires that the core thread has 100\% of the core context and can be sensitive to external processes affecting the utilization. It is, therefore, crucial that all unnecessary background applications and system functions are turned off. This utilization technique was compared with the output of the \textit{htop} utilization monitor and it revealed that the utilization was accurate to within $\pm1\%$.

The cluster frequencies are controlled using the \textit{Performance} frequency governor. The C application does not adjust the frequencies directly, it sets the maximum allowed frequency and the frequency governor then adjusts the frequency accordingly. 

\subsubsection{Thermal measurements}
\begin{figure}[!]
\centering
  \includegraphics[width=0.4\linewidth]{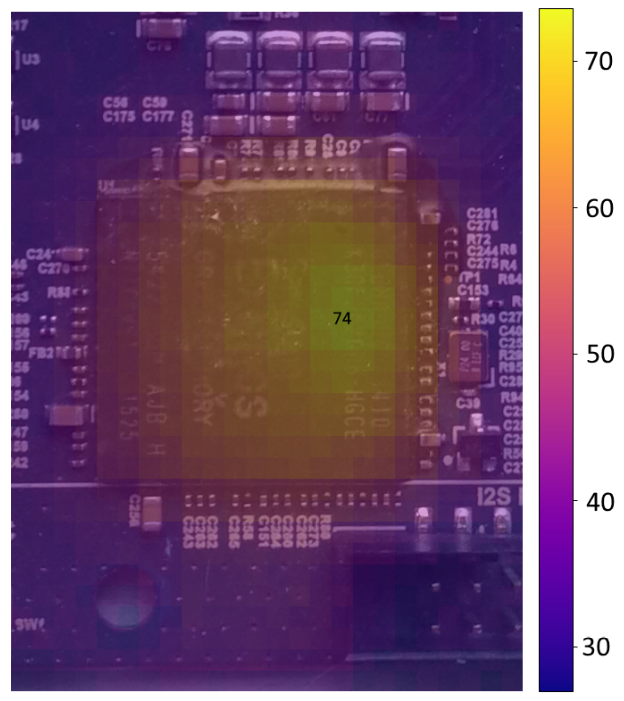}
  \caption{Heat map of the Odroid board.}
  \label{fig:heatmap}
\end{figure}

Due to the absence of a core temperature sensor for each core on the Odroid-XU4, a different temperature collection scheme had to be devised. The temperature can be quantified by measuring the emitted energy in the form of infrared radiation. Thermal imaging cameras are specialized sensors that measure the intensity of heat radiated by objects. For the setup in this work, the small thermal camera MLX90640 from Melexis has been used. It has a resolution of only 32 by 24 pixels. Therefore, the camera has been mounted close to the SoC of the Odroid-XU4, as can be seen in the right picture in Figure \ref{fig:setup}. This has been done in order to obtain a more accurate reading of the temperature across the surface or the SoC. The MLX90640 sensor has a temperature range of 40$^{\circ}$C to 300$^{\circ}$C and an accuracy of approximately $\pm 1 ^{\circ}$C. Figure \ref{fig:heatmap} shows a heat map of the Odroid board when the processor is running at 100\% utilization for all cores and the cluster frequencies are both set to $1500$ MHz. The region that dissipates the most heat is the region where the big cluster is located.

The measurements collected from thermal cameras are dependent on the emissivity of the objects that are being measured. In this experiment, the primary source of heat is expected to be the SoC chip. The emissivity value for a processor IC has been shown to be approximately 0.95 \cite{mashkov}. Because the SoC is the component expected to generate most of the heat on the board, this value is selected for the entire thermal image in this implementation.

\subsubsection{Cooling}
A drawback of using a thermal camera is that no heat sink can be mounted on top of the SoC. Thus, some external form of cooling had to be implemented in order for the processor not to overheat when running at higher clock speeds. The fan and Lego structure that can be seen in Figure \ref{fig:setup} provides sufficient directed cooling for the big cluster to be able to run at up to $1900$ MHz. In this implementation, the fan is constantly running at 100\% speed.

\subsubsection{Data collection}
A Raspberry Pi has been deployed as the control and data collection unit. It controls the experimental workloads and captures the thermal response. The data from the temperature sensor and the Odroid board were sampled 32 times per second. This sample rate was selected since it is the maximum sample rate for the thermal sensor.

The workload is in this work constant. Thus, the total number of possible configurations can be calculated using Equation \eqref{eq:configurations}, where $U$ is the number of utilization levels, $C$ is the number of cores, $f_b$ is the clock frequency of the big cluster and $f_l$ is the frequency of the little cluster. 

\begin{equation}
\label{eq:configurations}
N_c = U^C f_b f_l
\end{equation}

Each core has five different utilization levels, and the big and little clusters have ten and six discrete clock frequency levels, respectively. For the implementation in this work, this yields a total of approximately 23 million possible configurations of the heterogeneous SoC.

\section{Model creation}
We use system identification methods for creating the thermal dissipation model of the platform.

The power consumption has a nonlinear relationship with the core frequency, the core voltage and the core utilization. While the dynamic power dissipation is linearly dependent on the core utilization, the core utilization cannot, on its own, be used to describe it, as it is also dependent on the core frequency and voltage. Therefore, in this work, we selected the Polynomial N4SID approach for the system identification based on literature as best fitting for the non-linear nature of the head dissipation.

Polynomial N4SID \cite{jamaludin,VANOVERSCHEE199475} is a parametric approach based on the state-space identification method. We use the approach to construct a linear model of the system based on the direct relationship between the power dissipation of a processor and its thermal dissipation. A benefit of using an algorithm such as N4SID is that it is guaranteed to converge to the global minimum, non-iterative and numerically stable \cite{VANOVERSCHEE199475}.

The power consumption was shown to have a nonlinear relationship with the core frequency, the core voltage and the core utilization. While the dynamic power dissipation is linearly dependent on the core utilization, the core utilization cannot, on its own, be used to describe it, as it is also dependent on the core frequency and voltage. Therefore, a nonlinearity had to be introduced to approximate the power dissipation, in the form of new nonlinear regressors as polynomial combinations of the core frequency and core utilization. A linear state-space multi-input single-output model was then identified by the N4SID algorithm utilizing the nonlinear regressors as the input of the system. 

A state-space model with an input nonlinearity yields the set of equations shown in Equation \eqref{eq:kg-state-space}. The function $G(\cdot)$ maps the regressors $(u)$ to a new set of nonlinear regressor $(v)$.

\begin{equation}
\label{eq:kg-state-space}
\begin{aligned}
x(n+1)=& \ Ax(n)+Bv(n)+K\varepsilon(n)\\
y(n) =& \ Cx(n)+Dv(n)+\varepsilon(n) \\
v(n) =& \ G(u(n))
\end{aligned}
\end{equation}

If both the hidden states $x$ and the function mapping $G(\cdot)$ are unknown, this type of model becomes highly complex to identify. As no convergent identification methods such as N4SID exist for this nonlinear case, the simpler input-output models are commonly preferred \cite{nelles_nonlinear_2001}. 

\newcommand{\appropto}{\mathrel{\vcenter{
  \offinterlineskip\halign{\hfil$##$\cr
    \propto\cr\noalign{\kern2pt}\sim\cr\noalign{\kern-2pt}}}}}

The nonlinear input function for this model was derived from the relationship between the core frequency and the power consumption described previously. It was highlighted that the dynamic power consumption is proportional to $f V^2$. However, the voltage is not a quantity that is considered as it is directly affected by the core frequency. Therefore, the voltage part of the equation can be expressed in terms of its proportionality to the frequency. By using power regression, the approximate relationship was calculated to be $V \appropto \sqrt{f} $. Thus, the dynamic power can be expressed as $P_{dyn} \appropto f^{2}$. The cores in the little cluster have fewer clock frequency levels and even fewer voltage levels, but the same approximate relationship can be utilized for the regressors representing those cores. Furthermore, for the static portion of the power consumption, the same principle can also be applied. It was estimated to be approximately proportional to $f^{1.5}$. The core utilization is in this scenario expected to be directly proportional to the dynamic power consumption.

Using these approximate relationships as a basis, the polynomials were created as the product of the core utilization to a power of 0 or 1 and the core frequency to a power of between 1 and 3 in increments of 0.5. This was performed for each core and resulted in 58 new nonlinear regressors with a total of 68 regressors, including the original 10.

\begin{figure}[t!]
\centering
  \includegraphics[width=\linewidth]{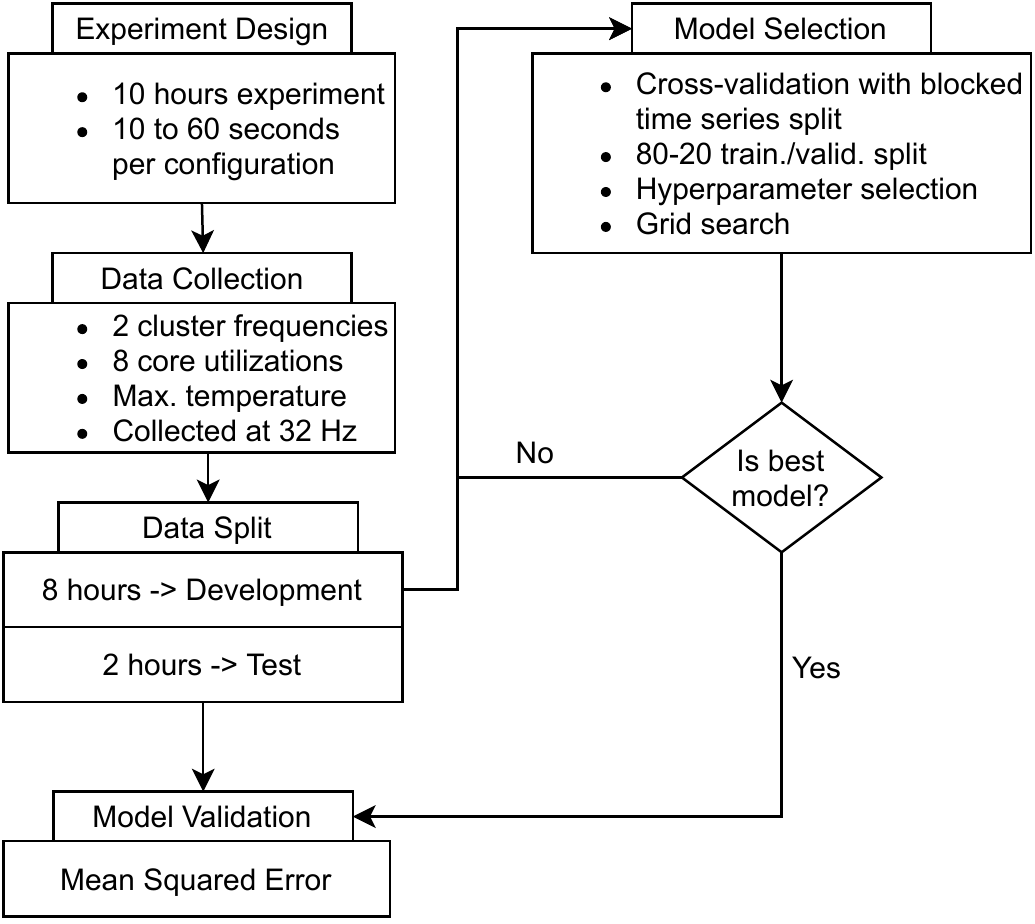}
  \caption{Flowchart of system identification procedure.}
  \label{fig:flowchart}
\end{figure}

The model creation approach has several steps as detailed in Figure~\ref{fig:flowchart}. First, we design the amount of data that needs to be collected for training the model, we collect the data, we choose a part of the data for training and a part for validation, and then we validate the accuracy of the model against the collected data. 

This approach was implemented in Python 3.7, using SIPPY, a system identification package developed at University of Pisa, Italy \cite{sippy}.

\subsection{Experiment design}
For training the model, 10 hours of data was collected from the Odroid computer board and thermal camera. The data set was created by executing a sequence of randomly selected configurations of the Odroid board using the stress application mentioned in Section \ref{sec:workload}. The configuration was changed after a random amount of time in the range of 10 to 60 seconds. Both the selection of configuration parameters (cluster frequencies and core utilization) and execution period followed a uniform distribution. Throughout the experiment, the ambient temperature was kept steady at 21$^{\circ} $C.

\subsection{Data Collection}
As mentioned earlier, the parameters that can be configured on the board are the frequencies of the two clusters and the utilization threshold for each core. Moreover, one temperature measurement is taken. Therefore, the data collected consists of 11 separate variables; a single regressand and 10 regressors.

The regressand is the maximum temperature measured by the thermal camera in degrees Celsius. The thermal image from the thermal sensor is captured at 32 Hz.

The first two regressors collected are the cluster frequency for each of two CPU clusters on the Odroid-XU4. For the implementation in this work, the values have been limited between $1000$ MHz and $1900$ MHz. The lower limit was included to restrict the number of configurations that will produce a very low thermal output. Including the lower frequencies would have made the data set rather imbalanced in favor of lower temperatures. The higher limit on the big cluster's operating frequency is imposed to assure that no core temperature reaches the thermal throttles point of 90$^{\circ}$C. 
The other eight regressors are the utilization for each core, with a granularity of 25\%. This allowed the utilization for each core to be selected at five discrete levels; 0, 25, 50, 75 and 100\%. All regressors were collected at 32 Hz to keep the sampling rate uniform across all collected variables.

\subsection{Data split}
The 10-hour-long data set was divided into two sets, a development set and a test set. The first 79\% of the data became the development set. This is the portion of the data that the models were trained on. The last 20\% of the data were chosen as the test set. This is the data set that the final prediction error was assessed upon and was not utilized for model training and selection. A small set of data corresponding to 1\% of the total data, lodged between the development and test sets, is omitted to ensure that there is no interference between the development set and the test set.

\subsection{Model selection}
The N4SID algorithm does not have many parameters that can be tuned. However, the model order can be viewed as a hyperparameter. In this implementation, the selection of nonlinear regressors can also be considered as hyperparameters. Additionally, the data preprocessing step also has to be considered. However, no preprocessing or filtering, except for resampling, was carried out for this experiment. The data for this modeling approach was resampled at 5 Hz. A rate that was established using grid search and cross-validation.

The optimal hyperparameters for the model have mainly been selected using grid search. Grid search is an exhaustive search method for testing which hyperparameter combinations that yield the best results for a model. Other search methods also exist, such as randomized search, which tests only a random subset of the hyperparameter combinations. The hyperparameters that yielded the lowest error on the validation set on average across all the folds were selected for the final model.

\begin{figure}[h!]
\centering
      \includegraphics[width=0.95\linewidth]{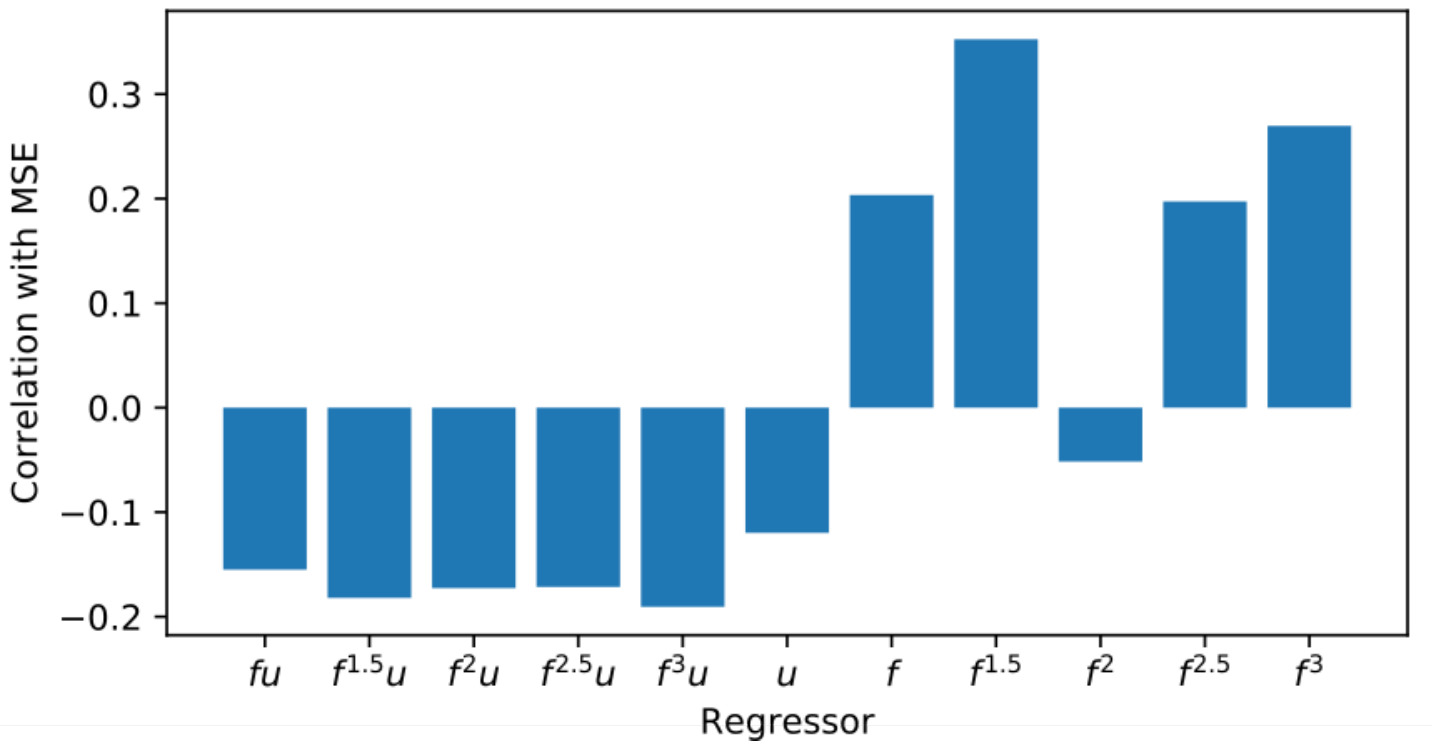}
  \caption{Correlation between regressor and MSE.}
  \label{fig:corr}
\end{figure}

Optimization of the utilized regressors was performed using correlation analysis and grid search. A phenomenon that was noticed during the early model selection phase was that using all the regressors led to some overfitting issues. Therefore, to reduce the number of regressors, randomized search cross-validation and correlation analysis was performed. The randomized search was performed with the 1-hour block length for 500 iterations. In each iteration, three random combinations of core frequency to a power between 1 and 3 and core utilization to a power of 0 or 1 were selected. The combinations were then applied to the regressors belonging to each core to create the new regressors. At the end of each iteration, the average MSE was measured. Using the results, a pair-wise correlation analysis was performed to detect each regressor's overall contribution to the error. Figure \ref{fig:corr} shows that most of the regressors with only a single frequency component showed a positive correlation. That is, they increased the error when they were utilized. Those that showed a negative correlation produced a decrease in the error when they were utilized. The regressors with a positive correlation were therefore removed from the regressor set.

Grid search and cross-validation were performed as an additional reduction step. During the grid search, the model order was set to 5 for all iterations. This was implemented to reduce computational time. The model order that produces the best performance was, however, expected to be higher than 5. An assumption was made, though, that a fifth-order model would be representative enough for this hyperparameter validation step. All permutations of the remaining regressors were tested and the best regressor configuration was saved. The best regressor set is shown in \eqref{eq:n4sidregessors}, where $f$ is core frequency, $u$ is core utilization and $i$ indicates the number of cores. 

\begin{equation}
\label{eq:n4sidregessors}
U_{nl} = [f^{1.5}u_i, f^{2}u_i, f^{3}u_i, u_i, f^2] , i=1..8
\end{equation}

The final number of regressors utilized in this approach is 34. Furthermore, these regressors were selected for implementation for both 1-hour and 6-hour block lengths. 

\begin{figure}[h!]
\centering
    \includegraphics[width=0.8\linewidth]{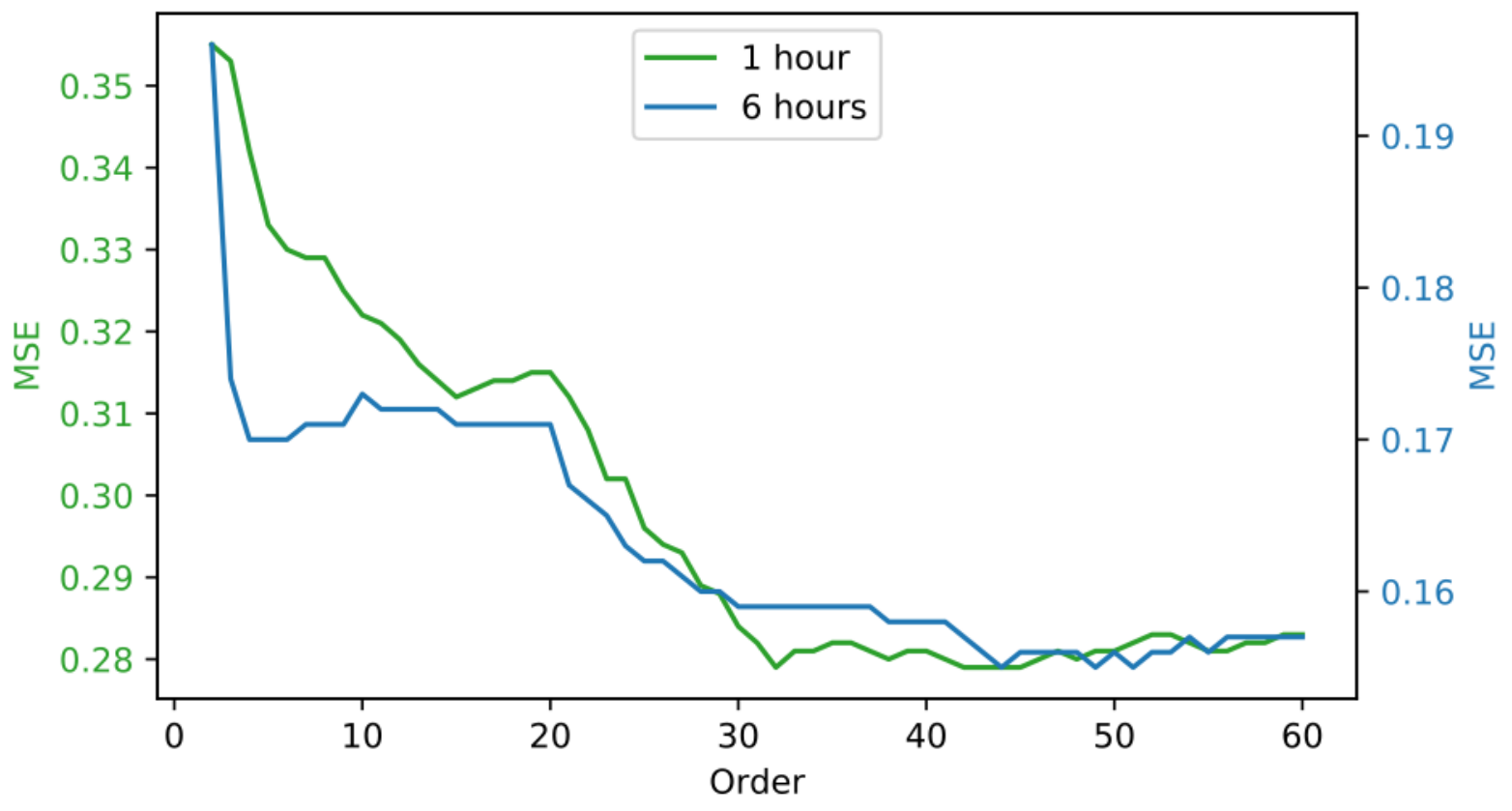}
  \caption{Validation error and model order.}
  \label{fig:orders}
\end{figure}

The optimal model order was estimated using the previously established combination of nonlinear regressors. The order of the state-space representation estimated by the N4SID algorithm was optimized using grid search and cross-validation. The average validation error was measured for orders between 2 and 60. Figure \ref{fig:orders} shows the model performance for each order.

The above figure shows that a model of lower order produces the best result for the 1-hour than for the 6-hour block length. The 1-hour model performed the best at 32 while the 6-hour model performed the best at 43.

Cross-validation is an umbrella term that refers to methods for analyzing how well a model will generalize to a set of independent data. A common practice in machine learning and data science is to utilize \textit{k}-fold cross-validation. The variable \textit{k} refers to how many subsets the data is divided into. In this type of cross-validation, the model is trained \textit{k} times, each time using a different subset as the validation set and the rest as the training set.

\begin{figure}[h!]
\centering
  \includegraphics[width=1.0\linewidth]{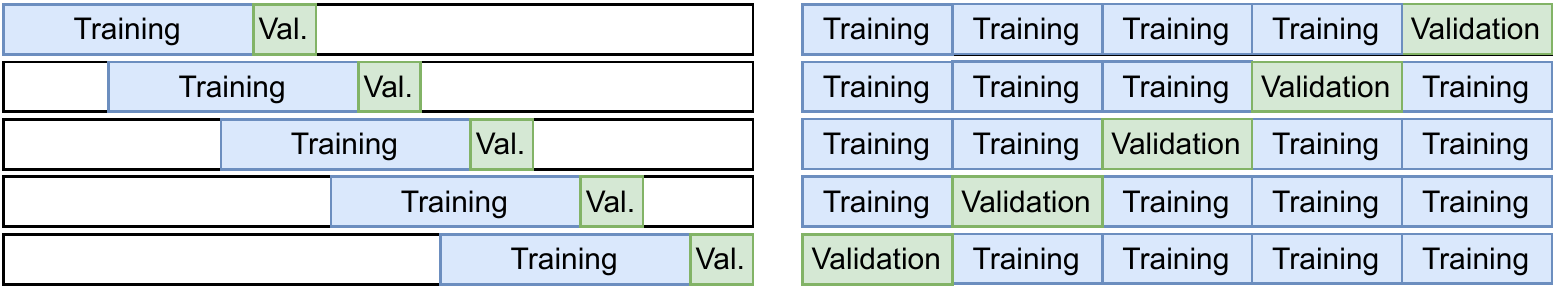}
  \caption{Comparison of \textit{k}-fold and blocked time series cross-validation}
  \label{fig:cv}
\end{figure}

In this work, since the data is a time series, each data point is dependent on the previous data points and the models are assessed for a specific length of training data, a special type of folding procedure called blocked time series split is utilized. Instead of dividing the data set into \textit{k} equal parts, like in regular k-fold cross-validation, the data is arranged into blocks of a specific length. Figure \ref{fig:cv} shows a 5-fold version of regular \textit{k}-fold cross-validation, to the right, and blocked time series cross-validation, to the left.

\begin{figure}[]
\centering
  \includegraphics[width=0.8\linewidth]{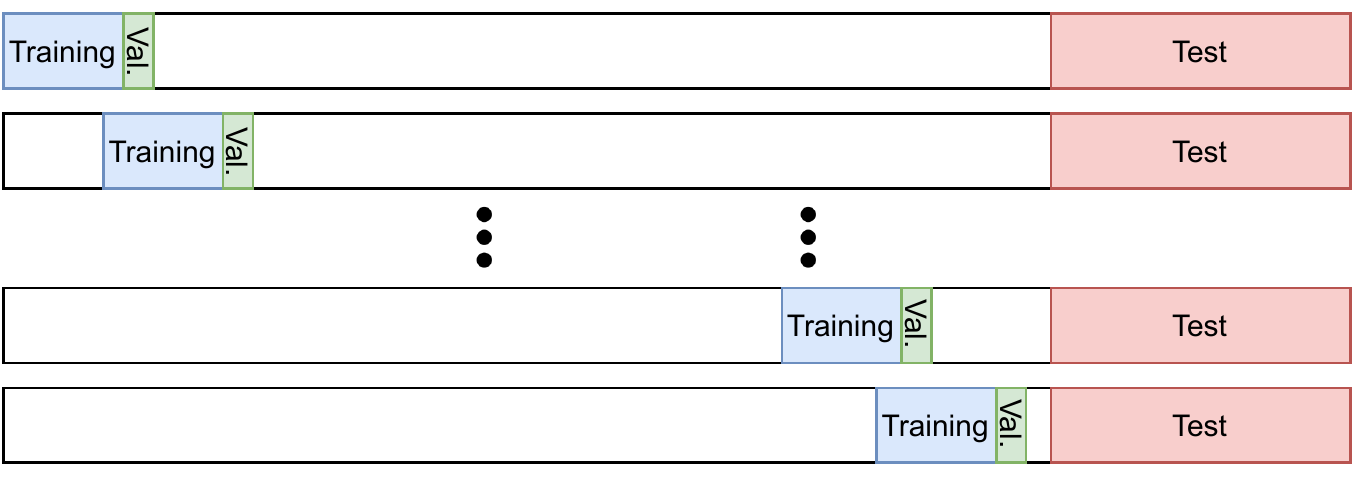}
  \caption{1-hour cross-validation procedure.}
  \label{fig:1hour}
\end{figure}

We evaluate the performance of the model with two data sets containing 1 hour and, respectively, 6 hour data. Just as with regular \textit{k}-fold cross-validation, some folds can have overlapping training data. However, the crucial thing is that no validation data is shared between folds. Therefore, an experiment length of 10 hours was chosen as it is the least amount of data that allows for dividing the data into multiple folds when evaluating the models' performance on 6 hours of data, while also having a small gap between the test set and any of the training or validation data.
In Figure \ref{fig:1hour}, the cross-validation procedure deployed when assessing the models' 1-hour performance is shown. On the development set, 10 blocks of 1 hour each with equal overlap, were selected. For each block in the cross-validation, the first 80\% of each block was utilized as the training set and the remaining 20\% as the validation set.


\begin{figure}[!]
\centering
  \includegraphics[width=0.8\linewidth]{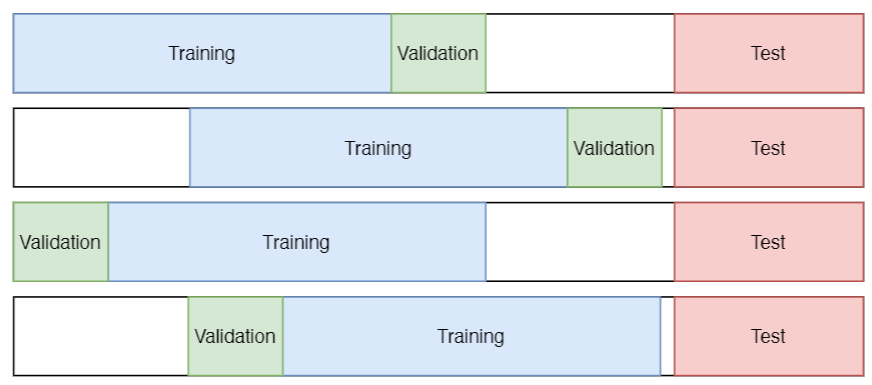}
  \caption{6-hour cross-validation procedure.}
  \label{fig:6hour}
\end{figure}

When assessing the model performance on 6 hours of training data, a slightly different method was utilized. In the 4-fold cross-validation, two blocks were created in the same manner as for 1 hour of data. The other two blocks were created by reversing the order of the validation and training data inside the block, as seen in Figure \ref{fig:6hour}. The validation set comprises the first 20\% and the training set the last 80\%. This reversed blocked time series split ensured that maximum diversity in the training and validation data is achieved.

The final prediction error was evaluated on the test set and the final performance metric was measured as the average MSE across the folds.

\begin{table}[]
    \setlength\tabcolsep{3pt}
    \centering
    \caption{MSE for the Polynomial N4SID method trained with 1 hour of data.}
    \label{tab:1hour-test}
    \begin{tabular}{|l|c|c|c|c|c|c|c|c|c|c|c|}   
    \hline
       & \multicolumn{10}{|c|}{\textbf{Folds}} & \\ \hline
        \textbf{} &1&2&3&4&5&6&7&8&9&10&\textbf{Avg} \\  \hline
        \textbf{MSE} &0.16	&0.15	&0.15	&0.16	&0.16	&0.14	&0.16	&0.16	&0.17	&0.14&\textbf{0.16}  \\ \hline
    \end{tabular}
\end{table}

\begin{table}[]
    \setlength\tabcolsep{3pt}
    \centering
    \caption{MSE for the Polynomial N4SID method trained with 6 hours of data.}
    \label{tab:6hour-test}
    \begin{tabular}{|l|c|c|c|c|c|}   
    \hline
       & \multicolumn{4}{|c|}{\textbf{Folds}} & \\ \hline
        \textbf{Method} &1&2&3&4&\textbf{Avg} \\  \hline
        \textbf{MSE} & 0.11 & 0.11 & 0.11 & 0.11 &  \textbf{0.11}  \\ \hline
    \end{tabular}
    
\end{table}

Table \ref{tab:1hour-test} shows the result for the model on the 1-hour block length and Table \ref{tab:6hour-test} shows the result for the model on the 6-hour block length.

\subsection{Model Evaluation} \label{cha:eval}
\begin{figure}[h!]
\centering
    \includegraphics[width=\linewidth]{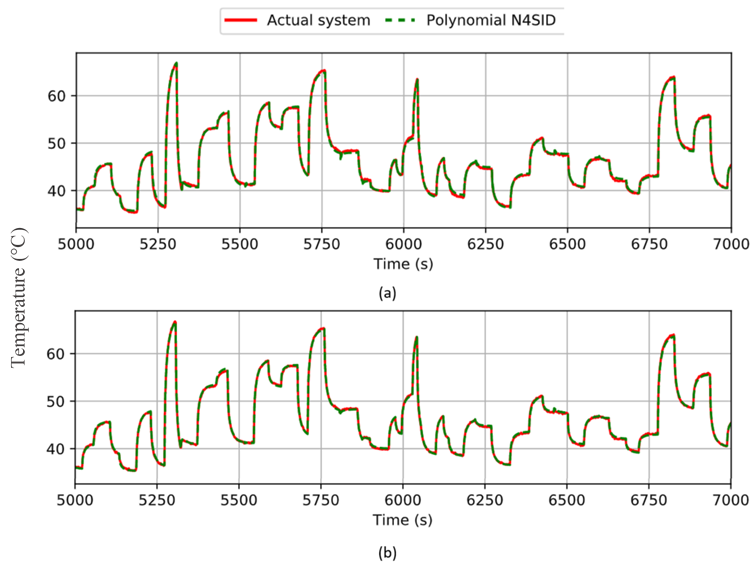}
  \caption{Model predictions vs measured test data for (a) 1-hour and (b) 6-hours models}
  \label{fig:1-6hourplot}
\end{figure}

The performance of the model was evaluated using a new data set based on a two hours long test session. The configuration parameters of the board were randomly changed every 10 to 60 seconds.This means that approximately 57 different board configurations were utilized in the 2000 second window. The results of the comparison between the model outputs and the physically measured temperature are shown in Figure \ref{fig:1-6hourplot}. 

Looking at the above figure, it can be seen that the Polynomial N4SID model produced a good approximation of the true measured temperature and it does not appear to have any particular problem areas or specific configurations that it struggles with.  


With respect to the performance of the model, the average training time, average prediction time and the number of parameters
for the N4SID-based model for both 1-hour and 6-hours of training data is shown in Table \ref{tab:6hour-comp}. This is in line with the training time complexity of the N4SID algorithm is $O(n^2)$ and the prediction time complexity is $O(n)$, where $n$ is the model order. 

\begin{table}[h]
     \small
    \setlength\tabcolsep{3pt}
    \centering
    \caption{Average training time, average prediction time and number of parameters for the 1-hour and 6-hour models}
    \label{tab:6hour-comp}
    \begin{tabular}{|l|c|c|c|}   
    \hline
        \textbf{Polynomial N4SID}& \makecell{Training \\ time (s)}&\makecell{Prediction \\ time (s)} & \makecell{ Number of \\ parameters} \\  \hline
        \textbf{1-hour} & 6	 & 	0.25  &  2144 \\ \hline   
        \textbf{6-hour} & 60	 & 1.34	  & 3354  \\ \hline 
    \end{tabular}
\end{table}

\section{Thermal exploration and validation}
Having the thermal prediction model constructed one can use it to quickly explore different SoC configurations without the use of the actual hardware platform. Depending on the precision and time available for the exploration, one can choose either the 1-hour or the 6-hour model if available. However, in a typical case, one model can be created and used.

As seen from Table \ref{tab:6hour-comp}, getting a prediction for one single configuration takes 0.25 and, respectively, 1.34 seconds, depending on the model used. In addition, for sequential experiments, the reconfiguration time of the platform and the cool-down time between the experiments are no longer needed. Therefore, the exploration can be done 400 times faster for the 1-hour model  and 74 times faster for the 6-hour model compared to the hardware measurement approach.

Also, using the models one can double check, with a high level of confidence,  whether a given configurations satisfies the thermal requirements of a system before deploying the application on hardware and with out requiring a setup for monitoring the temperature. 

\section{Limitations}
There are several limitations with our approach which could be addressed in the future work. The first is utilizing the ambient temperature and humidity as variables. In this work, they were not considered for technical reasons, as no climate-controlled environment was available. For real-world implementations this aspect is crucial and should be considered. Also the impact of active cooling was not taken into account.

Another important and perhaps also more complex aspect to examine in future work is the impact of the workload application. In this work, a static workload application was utilized. Many of the related works, however, suggest methods to represent a workload application as event counters, such as the number of operations of a certain type or the number of cache accesses that are performed. This would allow for a more generally applicable model as it would not be limit to a specific class of applications. 

A third aspect is related with multi-output systems. This work only considered the maximum temperature across the entire processor SoC, i.e., a MISO system. Predicting the thermal dissipation of individual parts of a CPU would allow for a more precise understanding of which parts of a heterogeneous SoC generate the most heat.

Last but not least we acknowledge that the generated prediction model could be biased by the experimental settings such as the used active cooling system and removal of the heat sink. We plan in future work to investigate the accuracy of the obtained model in a real-world setup.

\section{Conclusions}
We presented an approach for fast exploration and validation of thermal requirements of heterogeneous SoCs. The approach allows to speed up the evaluation and predictions of different hardware configurations without the need of the physical hardware and temperature measuring equipment. 
The approach poses a certain overhead in identifying the prediction model that provides good accuracy and which can vary from one platform to another depending on the characteristic of the platform characteristics and of the applications running on it. The time needed of collecting the training data set is most significant but depending on the level of accuracy needed, it can be shortened. However, once created, the model can be used for different evaluations tasks and with similar platforms and workloads. An example of such exploration task has been discussed in our previous research in \cite{Ivan9155898} where we used a discriminator neural network to perform the exploration of the configuration space of an Exynos 5422 SoC with respect to its power dissipation and performance.

Last but not least, we discussed the limitations of the approach presented in this paper which we plan to consider in future work.


\section*{Acknowledgment}

Part of this work was carried out with financial support from the Nordic Master programme  (contract NMP-2016/10169), ECSEL-JU MegaMArt2 project (grant agreement No 737494) and ECSEL-JU AIDOaRt project (grant agreement No 101007350). 

\bibliography{Bibliography}
\bibliographystyle{unsrt}

\end{document}